\documentclass[aps,twocolumn,prl]{revtex4}
\usepackage{epsfig}
\begin{document}

\title{The Dynamics of Crowd Disasters: An Empirical Study}
\author{Dirk Helbing, Anders Johansson}
\affiliation{Dresden University of Technology, Andreas-Schubert-Str. 23, 01062 Dresden, Germany}
\author{Habib Zein Al-Abideen}
\affiliation{Central Directorate of Development of Projects,
Minstry of Municipal and Rural Affairs, Riyadh, Kingdom of Saudi Arabia}
%\date{\today}
\begin{abstract}
Many observations in the dynamics of pedestrian crowds, including various 
self-organization phenomena, have been successfully described 
by simple many-particle models. For ethical reasons, however, there is a serious lack of experimental data regarding
crowd panic. Therefore, we have analyzed video recordings 
of the crowd disaster in Mina/Makkah during the Hajj in 1426H
on January 12, 2006. They reveal two subsequent, sudden transitions from laminar to stop-and-go 
and ``turbulent'' flows, which question many previous simulation models. 
While the transition from laminar to stop-and-go flows supports a recent
model of bottleneck flows [D. Helbing et al.,  Phys. Rev. Lett. 97, 168001 (2006)], the subsequent
transition to turbulent flow is not yet well understood.  It is responsible for sudden eruptions 
of pressure release comparable to earthquakes, which cause sudden displacements and the falling
and trampling of people. The insights of this study into the reasons for critical crowd conditions
are important for the organization of safer mass events. In particularly, they
allow one to understand where and when crowd accidents tend to occur. They have also 
led to organizational changes, which have ensured a safe Hajj in 1427H.
\end{abstract}

%Systems obeying scaling laws
%45.70.Ht}%Avalanches}
%47.27.Cn Transition to turbulence}
\maketitle

\section{Introduction}

The interest of physicists in pedestrian dynamics  dates back at least to the year 1995,
when a many-particle model was proposed to describe observed 
self-organization phenomena such as the directional segregation (``lane formation'') in pedestrian
counterstreams and oscillations of the passing direction at bottlenecks \cite{withMolnar}.
It took five more years until clogging effects and intermittent flows
in situations of crowd panic were discovered \cite{panic}. Since the year 2000, 
there is an avalanche of publications on pedestrians. 
This includes papers on other force models \cite{Yu} and
cellular automata models of pedestrian dynamics \cite{Nagatani,Schadschneider,Lattice}, addressing
counterflows \cite{Nagatani,Schadschneider,Weng}, 
the self-organized dynamics at intersecting flows \cite{Analytical}, 
capacity drops by interaction effects \cite{Jiang}, 
and the instability of pedestrian flows \cite{Nakayama}. 
Recent studies focus on the empirical or experimental study of pedestrian flows 
\cite{Lattice,Daamen,selforg,Isobe,Seyfried,Kretz1,Kretz2,Kretz3}
by means of video analysis \cite{Teknomo,Hoog}. 
\par
One of the most relevant and at the same time most challenging problems 
are panic stampedes, which are a serious concern during mass 
events \cite{Batty,Hughes,mahmassani,Still}. Despite huge numbers of
security forces and crowd control measures, hundreds of lives are lost in crowd disasters
each year. In this paper, we present a high-performance
video analysis of unique recordings of the Muslim pilgrimage
in Mina/Makkah, Saudi Arabia. It suggests that high-density
flows can turn ``turbulent'' and cause people to fall (see Sec.~\ref{trans}). The occuring eruptions of
pressure release bear analogies with earthquakes and are 
de facto uncontrollable. We will, however, identify variables that are helpful for 
an advance warning of critical crowd conditions. In our summary and outlook, 
we will indicate how our insights facilitated the organization of a safe Hajj in 1427H, i.e. 
during the stoning rituals between December 30, 2006, and January 1, 2007.

\section{Data Analysis and Fundamental Diagram}

While panic has recently been studied in animal 
experiments with mice \cite{mice} and ants \cite{ants}, there is still an evident lack of data on critical
conditions in human crowds. In homogeneous corridors and at very large openings,
unidirectional pedestrian flows are mostly assumed to move 
smoothly according to the "fluid-dynamic" flow-density relationship $Q(\rho) = \rho V(\rho)$, 
where $Q$ represents the flow, $\rho$ is the pedestrian density, and the average velocity $V$ 
is believed to go to zero at some maximum density as in traffic 
jams \cite{Fruin2,fund4,fund1,fund3,fund2,Seyfried} 
(see Fig.~1a). This formula is often used as basis for the dimensioning and design of pedestrian facilities, 
for safety and evacuation studies. Our video analysis of the Muslim pilgrimage
in Mina/Makkah shows, however, that this description needs corrections at extreme densities. 
In particular, it does not allow one to understand the ``turbulent'' 
dynamics causing serious trampling accidents in dense crowds trying to
move forward.
\par
We have evaluated unique video recordings of a 27.7m$\times$22.5m large area 
in front of the 44m wide entrance of the previous Jamarat Bridge, where upto
3 million Muslims perform the stoning ritual within 24 hours. On the 12th day of Hajj,
about 2/3 of the pilgrims executed lapidation even within 7 hours. 
With a new computer algorithm developed by us, 
we have extracted the positions $\vec{r}_i(t)$ and speeds $\vec{v}_i(t)$ of pedestrians $i$ 
as a function of time. This algorithm is based on the successive application of several digital
transformation, contrast enhancement, motion prediction and pattern recognition techniques
\cite{anders,preprint}. It has been calibrated with data obtained by manual evaluation and
tested in field experiments under well-known conditions, with an accuracy of about 95\%.
For the time period from 11:45am to 12:30 on January 12, 2006, the 
resulting dataset contains more than 30 million position-velocity pairs
in the evaluated {\em central} area of size 20m$\times$14m. We have restricted ourselves to the
evaluation of this area in order to avoid any boundary effects of the measurement algorithm. 
The resolution of 25 pixels per meter and 
8 frames per second allows one to determine even small average speeds by
calculating the mean value of a large sample of individual speed measurements.
\par
The data extracted from the videos allowed us to determine not only densities in larger areas,
but also local densities, speeds, and flows. The local density at place $\vec{r}=(x,y)$ and time $t$
was measured as
\begin{equation}
 \rho(\vec{r},t) =\sum_j f(\vec{r}_j(t)-\vec{r}) \, .
\label{dens1}
\end{equation}
$\vec{r}_j(t)$ are the positions of the pedestrians $j$ in the surrounding of $\vec{r}$ and 
\begin{equation}
f(\vec{r}_j(t)-\vec{r})  =  \frac{1}{\pi R^2} \exp[-\|\vec{r}_j(t) - \vec{r}\|^2/R^2]
\label{dens2}
\end{equation}
is a Gaussian distance-dependent weight function. 
$R$ is a measurement parameter. The greater $R$, the greater the effective measurement
radius, which is greater than $R$. It can be calculated that the weight of neighboring pedestrians located within
the area $A_R = \pi R^2$ of radius $R$ is 63\%. In another paper \cite{preprint}, we have shown that
the average of the local density values obtained with
formulas (\ref{dens1}) and (\ref{dens2}) agrees well with the actual average density $\varrho$. 
Moreover, the variance of the
local density measurements around the given, average density $\varrho$, 
goes down with larger values of $R$. 
In fact, for $R \rightarrow \infty$, all local density measurements result in the same
value, which corresponds exactly to the overall number $N_R$ of pedestrians, 
divided by the area $A_R = \pi R^2$ they are distributed in. 
The latter corresponds to the classical method of determining the
average (``global'') density $\varrho$. However,
it can also be determined by averaging over local density measurements $\rho_t^R(\vec{r})$, i.e.
\begin{equation}
 \varrho = \frac{N_R}{A_R}
 \approx \frac{1}{A_R}\int d^2r \, \rho_t^R(\vec{r}) 
 \approx \lim_{R \rightarrow \infty} \rho_t^R(\vec{r}_i) \, ,
\end{equation}
where the approximate equality becomes exact for $R\rightarrow \infty$. This is, because of
\begin{eqnarray}
& & \frac{1}{\pi R^2} \int \exp[-\|\vec{r}_j(t) - \vec{r}\|^2/R^2] \, d^2r  \nonumber \\
&=&  \frac{1}{\pi R^2}  \int\limits_0^{2\pi} \!\int\limits_0^\infty \!   
 \mbox{e}^{-r^2/R^2} r\, d\varphi \, dr = 1 \, ,
\end{eqnarray}
\begin{equation}
\lim_{R\rightarrow \infty} \exp[-\|\vec{r}_j(t) - \vec{r}_i(t)\|^2/R^2] = 1 \, ,
\end{equation}
and $N_R = \sum_j 1$.
\par
The local speeds have been defined via the weighted average 
\begin{equation}
 \vec{V}(\vec{r},t) = \frac{\sum_j \vec{v}_j f(\vec{r}_j(t)-\vec{r}) }
{\sum_j f(\vec{r}_j(t)-\vec{r})} \, ,  
\end{equation}
while flows have been determined according to the fluid-dynamic formula
\begin{equation}
 \vec{Q}(\vec{r},t) = \rho(\vec{r},t)\vec{V}(\vec{r},t) \, .
\end{equation}
In the fundamental diagram, one displays the absolute value of the flow as a function
of the density. For further aspects regarding the above definitions see Ref. \cite{RMP}.
\par
As the criticality in the crowd depends on the local conditions, we are interested in local
measurements with a small value of $R$, while many other measurements in the literature
present averages over larger areas. In the central recorded area studied by us, 
local densities reached values greater than 10 persons per square meter, if $R=1$m 
was chosen (Fig. 1a). The local densities
vary considerably (see Fig. 1c). As a rule of
thumb, the maximum local densities are twice as high as the average densities. 
When a local density of 6 persons per square meter is exceeded, 
the local flow decreases by a factor of 3 or more, so that the outflow drops 
significantly below the inflow (see Figs. 1b and 2b). This causes a higher and higher compression in the
crowd, until the local densities become critical in the end. 
While this seems to explain crushing accidents quite well \cite{Hughes,panic},
it is startling that the crowd accident on January 12, 2006 occured in a flat and practically open 
area without significant counterflows.
\par\begin{figure}[htbp]
\begin{center}
	\includegraphics[width=\columnwidth]{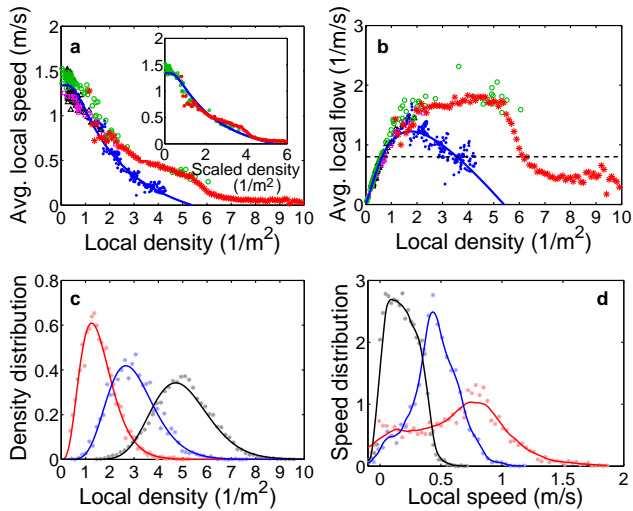} 
\end{center}
\caption[]{(Color Online)\textbf{a,} Average of the local speeds $V(\vec{r},t) = \| \vec{V}(\vec{r},t)\|$ 
as a function of the local density $\rho(\vec{r},t)$. Our own data points are for $R=1$m and shown as red stars. 
Other symbols correspond to data by Fruin \cite{Fruin} (black), Mori and Tsukaguchi \cite{fund1} (green), 
Polus {\em et al.} \cite{fund3} (purple), and Seyfried {\em et al.} \cite{Seyfried} (blue),
obtained with other measurement methods. The solid fit curve
is from Weidmann \cite{fund2}. Scaling the density with a factor 0.7 
(and the Mori data by a factor 0.6), our data become compatible with Weidmann's curve
(see inset), i.e. the different average projected body areas of people in different countries
are very important to consider \cite{Ph}.
Note, however, that the average local speed does not become zero at extreme densities. 
\textbf{b,} Average of the local flows $Q(\vec{r},t) = \rho(\vec{r},t)V(\vec{r},t)$ 
as a function of the local density $\rho(\vec{r},t)$. We have used the same symbols as in Fig. 1a. 
Note the second flow peak, i.e. the local maximum at 9 persons/m$^2$. 
\textbf{c,} Distribution of {\it local} densities $\rho$ for a given {\it average} density 
$\varrho$ (red: 1.6 persons/m$^2$,
blue: 3.0 persons/m$^2$, black: 5.0 persons/m$^2$). The Gamma distribution fits the histograms
with 50 bins well (solid lines).  
\textbf{d,} Distribution of local speeds for the same average densities $\varrho$ 
as in Fig. 1c (same colors for same densities). The distributions
deviate from the expected normal distributions, as many pilgrims are parts of large groups including
people of high age. Solid lines are smoothed fit curves serving as guides for the eye. Note that, at low
densities, a small percentage of pilgrims returns against the main flow direction.}
\end{figure}
Our video analysis revealed that, even at extreme densities,
the {\em average} local speeds and flows stayed finite, i.e. there was no level of crowdedness 
under which people completely stopped moving (Fig. 1a). This is in marked contrast to vehicle traffic \cite{NagataniReview,RMP}, 
where drivers keep some minimum safety distance and stop to avoid collisions. It also causes an unexpected, 
second flow maximum at 9 persons/m$^2$, which implies the possibility of alternating 
forward and backward moving shockwaves \cite{Colombo,PRL} with serious consequences for the resulting
crowd dynamics. Such shock waves cause safety hazards and 
have actually been observed \cite{shocks,Fruin,fund4,selforg}. 
However, a quantitative characterization of their properties and a satisfactory 
understanding of the underlying mechanisms are lacking.
Current models fail, as it is hardly conceivable that a 44m wide opening would 
cause intermittent outflows through arching effects \cite{panic}. 
Moreover, the notion of shockwaves is confusing, as it is mostly used for discontinuities in the 
density that are caused by random initial velocity or density variations, while we observe
emergent, self-generated waves (Fig. 2a). 

\section{Transition from Laminar to 
Stop-and-Go and ``Turbulent'' Flows}\label{trans}

\begin{figure}[htbp]
\begin{center}
\hspace*{-5mm}	\includegraphics[width=1.1\columnwidth]{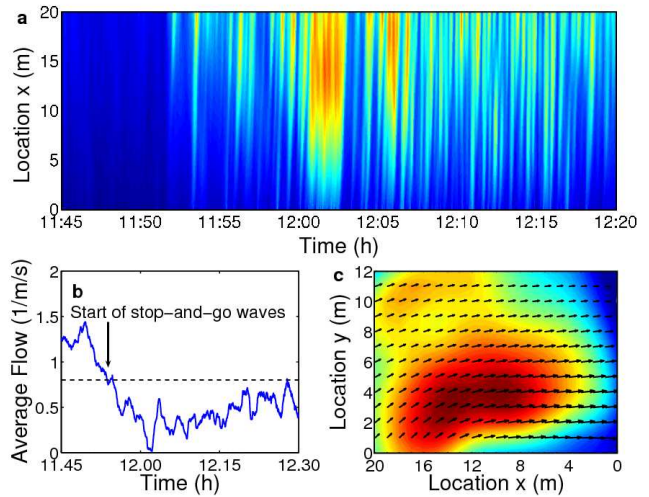}
\end{center} 
\caption[]{(Color Online)\textbf{a,} After a laminar phase, the density of pilgrims 
shows a sudden transition to stop-and-go waves around 11:53am. 
The densities shown were determined by Gaussian smoothing in space and time. Blue colors correspond
to low densities, yellow and orange colors reflect high values.
\textbf{b,} Average flow as a function of time. Note that the drop of the average pedestrian flow below 
a value of 0.8 persons per meter and second coincides with the occurence of stop-and-go waves, see Fig. 2a. 
\textbf{c,} Location-dependent velocity field $\vec{U}(\vec{r})= \langle \vec{V}(\vec{r},t)\rangle_{t}$
of pilgrim motion (where the bracket indicates an average over the index variable, i.e. over time). 
Arrows represent pedestrian speeds, averaged over the period
from 11:45am to 12:30am on January 12, 2006.  To avoid boundary effects, the 
evaluation focussed on the  20m$\times$14m central area of our video recordings.
The $x$-coordinate denotes the distance to the on-ramp of the Jamarat Bridge and 
points into the direction of its entrance.
One can clearly see the merging of pedestrians coming from different directions, 
which caused a bottleneck effect. The contour plot below the arrows represents the 
``pressure'' $P(\vec{r}) = \rho(\vec{r}) \mbox{Var}_{\vec{r}}(\vec{V})$, which we have 
defined as the average pedestrian density $\rho(\vec{r})$ %=\langle \rho(\vec{r},t)\rangle_t$ 
times the velocity variance $\mbox{Var}_{\vec{r}}(\vec{V})=\langle [V(\vec{r},t) - \vec{U}(\vec{r})]^2\rangle_t$ 
around the average velocity $U(\vec{r})$ \cite{Note3}. The dark red area represents the highest 
values of the ``pressure'', where the most violent dynamics occurred (see the crowd video in the
supplementary material \cite{supplement}). This is also the area where people stumbled and where the accident began.}
\end{figure} 
When viewing our video recordings in a 10 times accelerated fast-forward mode, we made 
some unexpected discoveries: As the average density increased, 
we observed a sudden transition from laminar to temporarily interrupted, longitudinally 
unstable flows around 11:53am (see the supplement \cite{supplement} and Fig.~2). The emergent, upstream moving stop-and-go waves 
persisted for more than 20 minutes and were not caused by the Hajj rituals. 
Around 12:19 and even higher densities, we found a sudden transition to irregular flows 
(see the supplement \cite{supplement} and Fig.~3), indicating a second instability. These irregular flows were characterized
by random, unintended displacements into all possible directions, which pushed people around. With 
a certain likelihood, this caused them to stumble. 
As the people behind were moved by the crowd as well and could not stop, fallen 
individuals were trampled, if they did not get back on their feet quickly enough. 
Tragically, the area of trampled people spread to larger and larger areas
in the course of time, as they became obstacles for others.
\par
Let us study these observations in more detail.
Due to the small acceleration times of pedestrians, the delay-based mechanism suggested to
describe stop-and-go waves in vehicle traffic \cite{NagataniReview,RMP} cannot be transfered to pedestrian crowds.
However, a recent theoretical approach suggests that intermittent flows at bottlenecks can also be generated by
coordination problems in bottleneck areas, causing an alternation between forward pedestrian motion
and backward gap propagation \cite{PRL}. This theory predicts a transition from smooth flows to stop-and-go patterns
when the inflow exceeds the outflow. In our videos of January 12, 2006, stop-and-go waves started 
in fact at the time when the outflow from the recorded area 
dropped below a value of 0.8 persons per meter and second (Fig. 2b), which supports this theory. 
\par
But how do we explain the second transition from stop-and-go 
to irregular flows (Fig. 3a, b), which occured at 12:19 (Fig. 3c)?
A closer look at our videos reveals that, at this time, people were so densely packed that they  
were moved involuntarily by the crowd. This
is reflected by random displacements into all
possible directions (see the crowd video in the supplement \cite{supplement}).
To distinguish these irregular flows
from laminar and stop-and-go flows and due to their visual appearance, 
we will refer to them as ``crowd turbulence''. 
\par
As in certain kinds of fluid flows, 
``turbulence'' in crowds results from a sequence of instabilities 
in the flow pattern. Additionally, we find the sharply peaked probability density function
of velocity increments 
\begin{equation}
 V_x^\tau = V_x(\vec{r},t+\tau) -V_x(\vec{r},t) \, ,
\end{equation}
which is typical for turbulence \cite{exchange}, if the time shift $\tau$ is small enough (Fig. 3d).
We also observe a power-law scaling of the displacements indicating self-similar behaviour (Fig. 3e). As we do not
observe large eddies, the similarity with {\em fluid} turbulence is limited, but there is still an analogy to 
turbulence at currency exchange markets \cite{exchange} (see Fig. 3d).
Instead of vortex cascades in turbulent fluids, we rather find a hierarchical fragmentation dynamics:
At extreme densities, individual motion is replaced by mass motion (Fig. 1a), but there is a stick-slip
instability which leads to ``rupture'' when the stress in the crowd becomes too large. That is,
the mass splits up into clusters of different sizes with strong velocity correlations {\em inside} and
distance-dependent correlations {\em between} the clusters. 
\par
\begin{figure}[htbp]
\begin{center}
 \includegraphics[width=1\columnwidth]{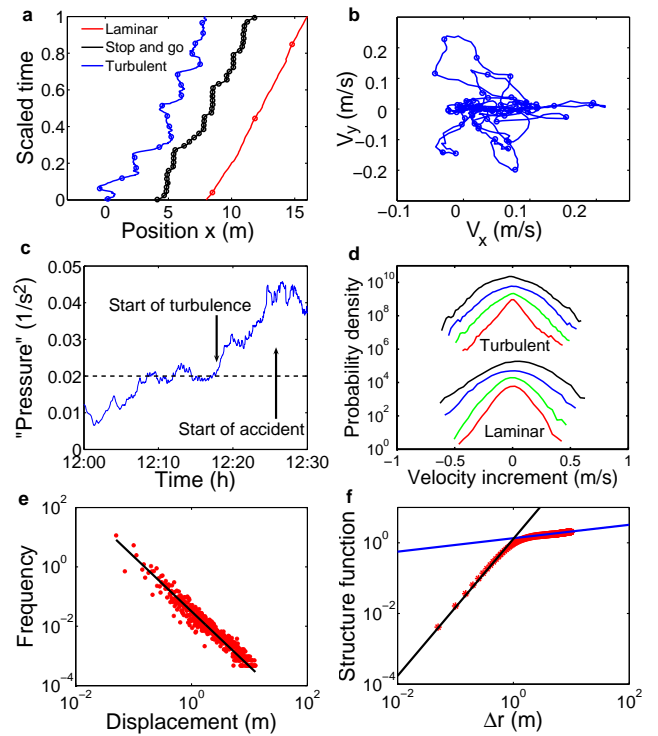} 
\end{center}
\caption[]{(Color Online)\textbf{a,} Representative trajectories in laminar flow, stop-and-go motion, and
``turbulent'' flow. Each trajectory extends over an $x$-range of 8 meters, while
the time required for this stretch was scaled to 1. To indicate the different speeds, symbols were included
in the curves every 5 seconds. While the laminar flow (red line) is fast and smooth, motion is temporarily interrupted
in stop-and-go flow (black line), and backward motion can occur in ``turbulent'' flows (blue line).
\textbf{b,} Example of the temporal evolution of the velocity components
$V_y(t)$ in $y$-direction and $V_x(t)$ in $x$-direction during ``turbulent'' crowd dynamics. 
A symbol is shown every second. One can clearly see the
irregular motion into all possible directions.
\textbf{c,} ``Pressure'' $P(t) = \rho(t) \mbox{Var}_{t}(\vec{V})$ as a function of time $t$, where
$\rho(t)$ is the spatial average of the density in the central recorded area 
and $\mbox{Var}_t(\vec{V})=\langle [V(\vec{r},t) - \langle V \rangle_{\vec{r}}]^2\rangle_{\vec{r}}$ 
is the velocity variance. (For the spatial dependence see Fig. 2c, for the spatio-temporal evolution, see
the video animation in the supplement \cite{supplement}.) The transition to ``turbulent'' crowd dynamics
(see the crowd video in the supplement \cite{supplement}) starts at 12:19, i.e. 
when the ``pressure'' exceeds the value 0.02/s$^2$.
The crowd accident began when the ``pressure'' reached its peak.
\textbf{d,} Probability density functions of the velocity increment 
$V_x^\tau = V_x(\vec{r},t+\tau) -V_x(\vec{r},t)$ in the laminar and the turbulent
regime, determined over many locations $\vec{r}$ for $R=\sqrt{10/\varrho}$ (see Fig. 1) and
$\tau =0.1$s (red curves), $\tau=1$s (green curves), $\tau=10$s (blue curves),
and $\tau = 100$s (black curves). For clarity of presentation, the curves are shifted in 
vertical direction. Note the non-parabolic, peaked curve for small values of $\tau$, which distinguishes
turbulent from laminar flows.
\textbf{e,} Distribution of displacements (i.e. location changes between subsequent stops, defined by
$\|\vec{V}(\vec{r},t)\| < 0.1$m/s). The double-logarithmic representation reveals a power law reminiscent
of a Gutenberg-Richter law for earthquake amplitudes. Here, the slope is 2.01$\pm$0.15. 
\textbf{f,} Double-logarithmic representation of the 
structure function $\langle \|\vec{V}(\vec{r}+\Delta \vec{r},t) - \vec{V}(\vec{r})\|^2\rangle_{\vec{r},t}$ 
of ``turbulent'' crowd motion, measuring the dependence of the relative speed on the distance $\Delta \vec{r}$.
As in fluids, the slope at small distances is 2, but the slope of 0.18$\pm$0.02 at large separations 
(in the so-called ``inertial regime'') differs from turbulent fluids due to the increased propulsion forces during ``crowd panics''.}
\end{figure}
``Crowd turbulence'' has further specific features.
Due to the physical contacts among people in extremely dense crowds, we expect
commonalities with granular media beyond the level of {\em analogy}
established in previous work \cite{TGF}. In fact, dense driven granular media 
may form density waves, while moving forward \cite{Peng}, and can display 
turbulent-like states \cite{turb1,turb2}. Moreover, under quasi-static conditions \cite{turb1}, 
force chains \cite{fragile} are building up, causing strong 
variations in the strengths and directions of local forces. As in 
earthquakes \cite{earthquake,earthquake2} this can lead to events 
of sudden, uncontrollable stress release with power-law distributed displacements. 
Such a power-law has also been discovered by our video-based crowd analysis (see Fig.~3d). 
\par
In contrast to purely density-based assessments, we suggest to quantity
the criticality of the situation in the crowd by the ``pressure''
\begin{equation}
 P(\vec{r},t) = \rho(\vec{r},t) \mbox{Var}_{\vec{r},t}(\vec{V}) \, ,
\end{equation}
i.e. the local pedestrian density times the local velocity variance $ \mbox{Var}_{\vec{r},t}(\vec{V})$.
Closeup video recordings show that, under ``turbulent'' conditions, the densities and 
resulting mechanical pressures are so unbearable that people try to escape 
the crowd and start pushing to gain space. This state, which is sometimes
called ``crowd panic'', is characterized by additional energy input in compressed areas, in contrast to
normal fluids or granular media. This causes particularly violent displacements in extremely dense crowds,
which are practically impossible to control even by large numbers of security forces and reflected by a
different scaling behaviour of the so-called structure function 
\begin{equation}
 S(\Delta \vec{r}) =\langle \|\vec{V}(\vec{r}+\Delta \vec{r},t) - \vec{V}(\vec{r})\|^2\rangle_{\vec{r},t}
\end{equation}
compared to fluids (Fig. 3f). 
Current simulation models of crowd panics fail to reproduce this ``turbulent'' dynamics, 
as they neglect that the propulsion force of people {\em increases} in areas of extreme densities. 

\section{Summary and Outlook}

In summary, even in extremely dense crowds with local densities upto 10
persons per square meter, the motion of the crowd is not entirely stopped.
This produces over-critical densities. The largest danger, however,
originates from the dramatically different crowd dynamics at high densities.
We have found two sudden transitions leading from laminar to stop-and-go flows and from there 
to ``turbulent'' crowd motion, which can trigger the trampling of people, in contrast to
previously discovered self-organisation phenomena in pedestrian crowds \cite{selforg}.
Stop-and-go waves start, when the density is high and the flow drops below a critical value
(Fig. 2a, b), while ``turbulent'' dynamics sets in with overcritical ``pressures'' in the 
crowd \cite{Note3}, see  Fig. 3c. The critical values depend on the choice of $R$ in the evaluation
of the local densities, speeds, and variances, see Eq.~(\ref{dens2}).
It is still an unresolved challenge to simulate {\em both} transitions, 
from laminar to stop-and-go {\em and}
``turbulent'' flows by a many-particle
model just by increasing the inflow to a bottleneck area.

\subsection{Practical Implications}

Due to the generality of these mechanisms, we expect that our findings are transferable to other 
mass gatherings. In fact, qualitatively similar conclusions can be drawn from video recordings at the same location
during the Hajj in the year 2005. In that year, the pressure did not reach so high values and
no accident occurred, but in 1997 and 1998 there were crowd disasters North of the ramp of the Jamarat
Bridge as well. Similar observations were reported by Fruin \cite{Fruin} from other places: 
``At occupancies of about 7 persons per square meter the crowd becomes almost a fluid mass.
Shock waves can be propagated through the mass, sufficient to ... 
propel them distances of 3 meters or more. ... People may be literally lifted 
out of their shoes, and have clothing torn off. 
Intense crowd pressures, exacerbated by anxiety, 
make it difficult to breathe, which may finally cause compressive asphyxia. 
The heat and the thermal insulation of surrounding bodies cause some to be weakened and faint. 
Access to those who fall is impossible. Removal of those in distress can only be accomplished 
by lifting them up and passing them overhead to the exterior of the crowd.''
This drastic picture visualizes the conditions in extremely dense crowds quite well,
but Fruin and others have not provided a scientific analysis and interpretation. 
\par
Turbulent waves are experienced in dozens of crowd-intensive events each year
all over the world \cite{Fruin}. Therefore, it is necessary to understand why, where and
when potentially critical situations occur.
Viewing real-time video recordings is not very suited to identify critical crowd conditions: 
While the average density rarely exceeds
values of 6 persons per square meter, the local densities can vary considerably 
due to dynamical patterns in the crowd (see Fig. 1c). Moreover, evaluating the local densities 
is not enough to identify the critical times and locations precisely, which also applies to an analysis of the 
velocity field \cite{supplement}. The decisive quantity is rather the variance of speeds, multiplied by 
the density, i.e. what we call the ``pressure'' \cite{Note3}. It allows one to identify critical locations (Fig. 2c) 
and times (Fig. 3c). There are even advance warning signs of critical crowd conditions:
The crowd accident on January 12, 2006 started about 10 minutes 
after ``turbulent'' crowd motion set in, i.e. after the ``pressure'' exceeded a value of 0.02/s$^2$
(Fig. 3c). Moreover, it occured more than 30 minutes after the average
flow dropped below a critical threshold (Fig. 2b), which can be identified by  
watching out for stop-and-go waves in accelerated surveillance videos (Fig. 2a).
Such advance warning signs of critical crowd conditions 
can be evaluated on-line by an automated video analysis system. In many cases, this can 
help one to gain time for corrective measures such as flow control, 
pressure relief strategies, or the separation of crowds into blocks to stop the propagation
of shockwaves \cite{supplement}. 
Such anticipative crowd control could certainly increase the level of safety during 
future mass events.

\subsection{Implications for the Hajj in 1427H}

Based on our insights in the reasons for the accidents during the Hajj in 1426H, we
have recommended many improvements. In the following, we mention only the most important
changes in the organization of the Hajj in 1427H: 
\begin{itemize}
\item The stoning capacity of the Jamarahs and, thereby, the flow capacity 
of the different levels of the Jamarat Bridge
was improved by an elongated, elliptical shape, as suggested by Dr. Keith Still.
\item On the plaza around the Jamarat Bridge, no opportunity for accumulation of the crowd
was given. 
\item An automated counting of pilgrims using the signals of the surveillance cameras
was implemented to have all the time a reliable overview over the densities and capacity utilizations in 
critical places of the system.
\item Complementary, a new plaza design allowed the General Security
to easily balance the flows between the ground floor of the current construction stage 
of the new Jamarat Bridge and the Northern and Southern ramps of the first floor, in order to avoid overloading 
and a breakdown of the flow.
\item The two-way operation of the street system and the Jamarat plaza was replaced by
a one-way operation in order to avoid obstructions and problems by counterflows.
\item A systematically optimized scheduling and routing program was applied in order
to reach a homogeneous distribution of registered pilgrims in space and time.
\item Squatters were removed from the streets in order to avoid bottleneck situations.
\end{itemize}
These and some further changes (which will be described in detail in some other publications)
ultimately reached comfortable and continuous flows conditions
and a safe Hajj in 1427H, although the situation was expected to be particularly critical due to
a lack of experience with the implemented changes and due to the significantly increased number of
pilgrims in 1427H. As the new Jamarat Bridge will be expanded in 2007 by additional floors, it will have
a greater capacity in 1428H and imply the need for changes in the organizational concepts to
avoid bottlenecks in other places. This will pose new challenges for crowd researchers,
experts, and authorities in the future. 

\subsection*{Acknowledgements} 

The authors are grateful to the German Research Foundation for funding 
(DFG project He 2789/7-1), to the Ministry of Municipal and Rural Affairs
for providing data and organisational support, and to its minister, H.R.H., for 
facilitating this scientific collaboration. They also thank Salim Al-Bosta and the Stesa staff 
for spending many hours positioning the cameras, converting the video recordings, and the
great cooperation and support. Finally, D.H. appreciates the insightful discussions with
various colleagues and the phantastic collaboration with Dirk Serwill, Knut Haase, Erfan Qasimi
and many others, who have contributed to the progress of this project. He congratulates
the Saudi authorities to their great efforts and the successful implementation.

\end{document}